\documentclass[preprint,showpacs,preprintnumbers,amsmath,amssymb]{revtex4}
\usepackage{graphics,epsfig,subfigure}
\usepackage{color}
\usepackage[linktocpage]{hyperref}

\begin{document}
\renewcommand{\baselinestretch}{1.3}
\newcommand\beq{\begin{equation}}
\newcommand\eeq{\end{equation}}
\newcommand\beqn{\begin{eqnarray}}
\newcommand\eeqn{\end{eqnarray}}
\newcommand\nn{\nonumber}
\newcommand\fc{\frac}
\newcommand\lt{\left}
\newcommand\rt{\right}
\newcommand\pt{\partial}
\allowdisplaybreaks

\title{A braneworld model in a massive gravity} 
\author{
Ke Yang$^a$\footnote{keyang@swu.edu.cn}, 
Shi-Fa Guo$^{b}$\footnote{shifaguo97@gmail.com},
Bao-Min Gu$^{c,d}$\footnote{gubm@ncu.edu.cn, corresponding author}
}

\affiliation{
$^{a}$School of Physical Science and Technology, Southwest University, Chongqing 400715, China\\
$^{b}$Center for Joint Quantum Studies and Department of Physics, School of Science, Tianjin University, Tianjin 300350, China\\
$^{c}$Department of Physics, Nanchang University, Nanchang 330031, China\\
$^{d}$Center for Relativistic Astrophysics and High Energy Physics, Nanchang University, Nanchang 330031, China}

\begin{abstract}

A Randall-Sundrum-like braneworld scenario is constructed in a 5D extension of the Lorentz-violating massive gravity. The gauge hierarchy problem is solved in current model. The linear perturbations are calculated, and it is found that the tensor and vector perturbations are robust and free from the ghost and tachyonic instabilities, however, the scalar perturbation is a ghost filed. After Kaluza-Klein reduction, all the tensor, vector and scalar modes are massive and possess the mass splitting of order of TeV in their respective mass spectra. The massive ground states of tensor and scalar modes propagate only along the brane, however, the vector ground state is absent in the mass spectrum.  By introducing the Goldberger-Wise mechanism to stabilize the extra dimension, the 4D effective theory on the brane includes a nearly massless graviton plus three towers of very massive spin-2,  spin-1 and ghost spin-0 particles.

\end{abstract}











\maketitle



\section{Introduction}

Since Kaluza and Klein proposed the idea that our fundamental spacetime may be more than four dimensions in 1920s \cite{Kaluza1921,Klein1926}, extra dimension theories have received much attention from theoretical physicists. In Kaluza-Klein (KK) theory, the extra dimension is compactified as a circle with the radius of Planck length $l_{Pl} \sim 10^{-35}$m, and this strategy of compactifying extra dimensions dominates higher-dimensional unified physics \cite{Overduin1997}. Inspired by the string theory, the braneworld scenario provides another way to hide the extra dimensions.  In this scenario, our universe is a brane embedded in a higher-dimensional bulk. All the Standard Model particles are trapped on the brane and only the graviton can propagate into the extra dimensions. 

In 1999,  Randall and Sundrum proposed a well-known braneworld model \cite{Randall1999}, which is now dubbed RS1 model. In this model, the bulk is a slice of a five-dimensional anti-de Sitter (AdS) spacetime, and there is a brane at each boundary of the slice. In order to maintain a flat four-dimensional spacetime on the brane, the fine-tuning condition requires that the tension of the ultraviolet (UV) brane be positive while the tension of the infrared (IR) brane be negative. Our universe lives on the IR brane, so owing to the exponentially warped extra dimension, a fundamental electroweak scale of the order of Planck scale ($M_{Pl}\sim 10^{16}$TeV) is redshifted to the observed effective electroweak scale ($\sim$246GeV). The RS1 model provides a natural way to solve the gauge hierarchy problem without introducing a new large hierarchy, so it draws lots of attention in the last two decades, see, e.g., Refs.~\cite{Rubakov2001,Csaki2004,Maartens2010,Gherghetta2010,Csaki1999,Cline1999,Goldberger1999a,Rattazzi2001,Kisselev2016,Blanke2018} for introduction and recent works. 

The RS1 model is based on the 5D general relativity (GR), where the 5D graviton is massless and possesses 5 degrees of freedom (DOF), therefore, there is a massless 4D graviton, a massless scalar radion, and a tower of massive 4D graviton in the effective 4D theory on the brane \cite{Randall1999}. In the braneworld scenario, the effective 4D theory on the brane is related to the DOF of the high-dimensional gravitational theory. So it is natural to ask what is the difference between the 4D effective theories on the brane if the higher-dimensional graviton possesses different numbers of DOF. One simple exploration is to replace GR with the massive gravity theory, where the higher-dimensional massive graviton possesses more numbers of DOF. 

Massive gravity theory is a generalization of GR by endowing the graviton with a nonzero mass. The first pioneering work was done by Fierz and Pauli in 1939 by adding non-derivative interaction terms, known as the Fierz-Pauli mass term, into the linearized level of GR \cite{Fierz1939}. However, the theory does not reduce to GR in the massless limit, so it suffers from a discontinuity problem, known as van Dam-Veltman-Zakharov discontinuity \cite{Dam1970,Zakharov1970,VanNieuwenhuizen1973}. This problem can be alleviated by generalizing the linear Fierz-Pauli mass term to nonlinear terms \cite{Vainshtein1972}. However, the nonlinear terms will generate a higher derivative term and lead to the Boulware-Deser ghost instability \cite{Boulware1972}. A breakthrough was achieved by de Rham, Gabadadze, and Tolley about a decade ago \cite{Rham2010a,Rham2011}. They constructed a nonlinear massive gravity theory with the field equation at most second order in time derivatives by getting rid of ghosts for all nonlinear self-interactions of the longitudinal component in the decoupling limit \cite{Hassan2012,Hassan2012a}. This massive gravity theory is now dubbed dRGT gravity. However, it is found that the dRGT gravity may suffer from other pathologies. In Ref.~\cite{Deser2013}, Deser and Waldron argued that the theory admits second order superluminal shock wave solutions, and the acausal characteristic can arise for any choice of background. A detailed and correct analysis of characteristic equations was performed by Izumi and Ong \cite{Izumi2013a}, and they showed that the theory does admit a well-posed Cauchy problem. Such acausality problem were further discussed in  Refs.~\cite{Deser2013a,Deser2014a,Hassan2018}. The problem of lacking the stable and realistic Friedmann-Lema\text{\^ \i}tre-Robertson-Walker cosmological solutions was discussed in Refs.~\cite{DAmico2011,Gumrukcuoglu2011,Gumrukcuoglu2012,DeFelice2012a,DeFelice2013}. 

On the other hand, although the Lorentz invariance is widely regarded as a fundamental symmetry of nature, there are still some hints that the Lorentz invariance may be broken at high energies in quantum gravity \cite{Kostelecky1989,AmelinoCamelia2003}.  Especially, the aforementioned standard problems of Lorentz-invariant massive gravity theories may be avoided by switching to the Lorentz-violating massive gravity theory, see Refs.~\cite{Rubakov2004,Dubovsky2004,Dubovsky2005,Blas2009,Lin2013} for examples.  A simple example of Lorentz-violating massive gravities was proposed by Lin in Ref.~\cite{Lin2014}, where the theory is built with the Einstein-Hilbert action plus 3 canonical massless scalar fields. The spatial condensation of the scalar fields breaks spatial diffeomorphisms, which gives rise to 3 Goldstone excitations. After ``eating" the 3 Goldstone bosons in the unitary gauge, the graviton gets weight and possesses 5 well-behaved DOF. It was found that the theory could remove the infrared divergence of the inflationary loop diagram owing to the graviton mass \cite{Lin2014}.  For a recent review on massive gravity theories see Refs.~\cite{Hinterbichler2012,Rham2014,Rham2017} and the references therein.

The higher-dimensional scenario has a close relation with massive gravities, as it naturally incorporates a tower of non-pathological massive spin-2 gravitons in the 4D graviton spectrum. Nevertheless, they are usually not a purely massive gravity theories due to the presence of a massless graviton associated with the unbroken 4D diffeomorphisms. A more interesting theory is the Dvali-Gabadadze-Porrati model, whose graviton looks like a resonance and effectively acquires a soft mass \cite{Dvali2000,Dvali2000a}. In Ref.~\cite{Chacko2004}, the authors considered the RS1 model with a localized Fierz-Pauli mass term for the graviton on the IR brane, which reproduces GR for observers localized on the UV brane at distances smaller than the IR scale but realizes a massive gravity at longer distance with a ghost radion. The embedding of a 4D dRGT massive gravity into a 5D dRGT massive gravity by resorting to the single-brane RS2 model was studied by Gabadadze in Ref.~\cite{Gabadadze2017}, where the value of strong scale $\Lambda_3$ can be significantly increased. More relevant works see Refs.~\cite{Gabadadze2004,Kakushadze2014,Gabadadze2018,Kaloper2019} for examples. 

In this work, we are interested in generating the massive gravity with spatial condensation into a 5D spacetime and studying the braneworld scenario in the massive gravity. The background spacetime in the spatial condensation scenario exhibits invariance under translation and 3D rotations. In order to achieve a Minkowski flat brane model, we require that the background spacetime is invariant under the 4D Poincar\'e transformation in the higher dimensional generalization. The condensation of 4 background scalars spontaneously breaks the 4D diffeomorphisms, and generates 4 Goldstone excitations associated with the broken symmetries. Consequently, by ``eating" the 4 Goldstone excitations in the unitary gauge, the 5D massless spin-2 graviton with 5 DOF becomes massive and possesses 9 DOF on the spectrum. 

 The layout of the paper is as follows: In Sect.~\ref{Model}, the toy model is built in the 5D massive theory. In Sect.~\ref{Perturbation}, the full linear perturbations have been considered. In Sect.~\ref{Hierarchy}, the mass spectra of KK particles have been calculated and the resolution of the gauge hierarchy problem has been discussed as an application of the model. In Sect.~\ref{Radius_stabilization}, the Goldberger-Wise (GW) mechanism has been used to stabilize the radius of the extra dimension. Finally, brief conclusions are presented. Throughout the paper, the small Latin letters $a, b, \cdots=0,1,2,3,$ label the group indices of the internal metric of scalar fields, and the capital Latin letters $A, B, \cdots=0,1,2,3,5,$ and Greek letters $\mu,\nu, \cdots=0,1,2,3,$ label the 5D and 4D spacetime indices, respectively.

\section{Model building}\label{Model}

We start from a 5D extension of the Lorentz-violating massive gravity \cite{Lin2014}, and the action is given by a 5D Einstein-Hilbert term plus 4 canonical massless scalar fields, i.e.,
\beqn
S&=&M^3\int{d^5x}\sqrt{-g}\lt[\fc{R}{2}-\fc{1}{2}m^2g^{MN}\pt_M\phi^a\pt_N\phi^b\eta_{ab} -\Lambda\rt]-\int d^4 x \sqrt{-g_{\text{I}}} V_{\text{I}}\nn\\
&&-\int d^4 x \sqrt{-g_{\text{II}}} V_{\text{II}},
\label{Main_Action}
\eeqn
where $M$ is the 5D fundamental gravity scale, $\Lambda$ is the 5D cosmological constant, $V_{I}$ and $V_{II}$ represent the brane tensions. The parameter $m$ is proportional to the mass of the 5D graviton, and this relationship will become evident in Sect.~\ref{Hierarchy}. Here, the internal metric $\eta_{ab}$ for the scalar fields is selected to be the Minkowski metric. This choice guarantees the preservation of 4D Poincar\'e symmetry, allowing for the attainment of a Minkowski flat brane model. However, it is evident that the utilization of the Minkowski metric results in one of the bulk scalars, namely $\phi^0$, having a kinetic term with the wrong sign. Consequently, this gives rise to a ghost scalar mode in the linear perturbation theory, which will be clearly observed in Sect.~\ref{Perturbation}. 

The background solution spontaneously breaks the 5D Lorentz invariance when the scalar fields are fixed to their vacuum values 
\beq
\langle\phi^a\rangle=\delta_\mu^a x^\mu, \label{Scalar_vacuum}
\eeq 
with $x^\mu$ the coordinates of 4D spacetime. Thus, the condensation spontaneously generates a preferred 4D frame.

In order to investigate a RS1 model-like braneworld solution, the metric ansatz  preserving the 4D Poincar\'e invariance reads
\beq
ds^{2}_{5}=a^2(y){\eta}_{\mu\nu}dx^{\mu}dx^{\nu}+dy^2,
\label{Brane_Metric}
\eeq
where $a(y)$ is the warp factor and $y \in [-y_\pi, y_\pi]$ denotes an $S^1/Z_2$ orbifold extra dimension. The UV brane with tension $V_\text{I}$ locates at $y=0$ and the IR brane with tension $V_{\text{II}}$ locates at $y=y_\pi$.

The field equations are obtained by varying the action (\ref{Main_Action}) with respect to the metric $g^{MN}$, 
\beqn
G_{MN}&=&m^2 \Big(\partial_M \phi ^a \partial_N  \phi ^a-\frac{1}{2}  g_{MN} \partial^K \phi ^a \partial_K \phi ^a\Big)-\Lambda  g_{MN}-\frac{\sqrt{-g_{\text{I}}} }{\sqrt{-g}}\fc{V_{\text{I}}}{M^3 }g_{\text{I}\mu \nu} \delta ^{\mu }_M \delta ^{\nu }_N \delta  \left(y\right)\nn\\
&&-\frac{\sqrt{-g_{\text{II}}} }{ \sqrt{-g}}\fc{V_{\text{II}}}{M^3}g_{\text{II}\mu \nu } \delta ^{\mu }_M \delta ^{\nu }_N \delta  \left(y-y_\pi\right).
\eeqn
Further, with the ansatzes of braneworld metric and scalar field condensation, the field equations are written explicitly as
\beqn
3\lt(H'+2H^2 \rt)\!&=\!&-\fc{m^2}{a^2}-\Lambda-\fc{ V_\text{I} \delta\left(y\right)+ V_{\text{II}} \delta\left(y\!-\!y_{\text{b}}\right)}{M^3},\label{EoMII_1}\\
6H^2\!&=&\!-\fc{2m^2}{a^2}-\Lambda, \label{EoMII_2}
\eeqn
where $H\equiv a'/a$ and the prime denotes the derivative with respect to the extra dimension coordinate $y$. 

\begin{figure}[htb]
\begin{center}
\includegraphics[width=8cm]{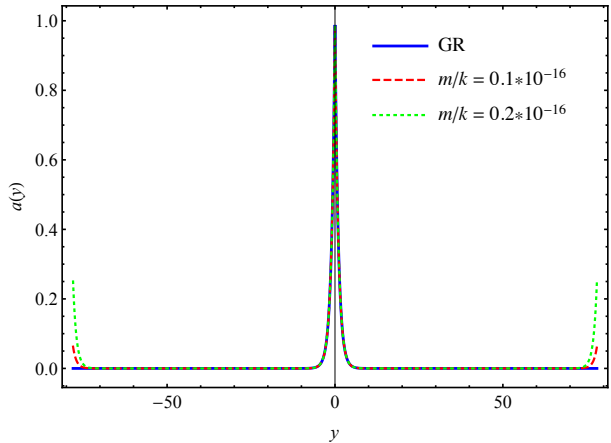}
\end{center}
\caption{The profiles of the warp factor $a(y)$ with respect to different mass parameters.}
\label{Fig_Warp_factor}
\end{figure}

The field equations simply give the following solution
\beq
a(y)=e^{-k|y|}+\epsilon^2e^{k|y|}, 
\label{Warp_factor}
\eeq
where $\epsilon\equiv\fc{m}{2\sqrt{3}k}$, $k^2\equiv-\Lambda/6$, and $|y|$ represents the absolute value of $y$ in order to be consistent with the $Z_2$ symmetry. Since the minimum value of warp factor is $2\epsilon$, the 5D graviton has to be light enough, i.e., $\epsilon<a(y_\pi)\sim10^{-16}$, to generate enough redshift for the fundamental electroweak scale.  As illustrated in Fig.~\ref{Fig_Warp_factor},  the warp factor is different from that of RS model only far from the center for a tiny graviton mass, $m/k\ll1$.

After matching the delta functions in \eqref{EoMII_1}, we have the fine-tuning conditions,
\beqn
V_I&=&6 k M^3 \left[1-\frac{2 \epsilon^2}{1+\epsilon^2}\right],\\
V_{II}&=&6 k M^3 \left[1-\frac{2}{1+\epsilon^2 e^{2 k y_\pi}}\right].
\eeqn

In the case of $e^{-k y_\pi}>\epsilon$, the exponential decay term of  warp factor \eqref{Warp_factor} is dominant in the whole bulk. It is clear that $V_I>0$ and $V_{II}<0$ in this case, namely, there is a positive tension brane located at the origin and a negative tension brane at $y_\pi$. This is the model we will focus on in the rest of the paper. Especially, in the limit $\epsilon e^{k y_\pi}\ll  1$, the warp factor \eqref{Warp_factor} approaches the exponential decay form of RS model. Correspondingly, the fine-turning conditions reduce to the case of RS model as well,
\beq
V_I\approx-V_{II}\approx6 k M^3, ~~~ \Lambda=-6k^2.
\eeq

Besides, there is another interesting case that both brane tensions are positive in case of  $e^{-k|y_\pi|}<\epsilon<1$. It would generate a new brane configuration distinct from the RS1 model. Especially, the scenario that our universe is confined on the positive tension brane may potentially provide a possible model with better properties \cite{Csaki1999,Shiromizu2000,Yang2012a}. This model will be studied in detail in our another work.

\section{Linear perturbations}\label{Perturbation}

In order to investigate the stability of the model, we consider the full linear perturbations against the background. The perturbed metric is written as
\beq
ds^2=\lt(g_{MN}+h_{MN}\rt)dx^Mdx^N,
\label{Metric_NC}
\eeq
where $h_{MN}$ represents the perturbations against the background metric $g_{MN}$ given in Eq.~\eqref{Brane_Metric}. Due to the 4D Lorentz symmetry of our background spacetime, it is convenient to decompose the perturbed metric into the scalar, transverse vector and transverse-traceless tensor modes, and to rewrite it by
\beqn
h_{55}\!&=\!&-2\xi,\label{h55}\\
h_{\mu 5}\!&=\!&-a \lt(S_{\mu}+\pt_\mu \beta \rt),\label{hmu5}\\
h_{\mu\nu}\!&=&\!a^2\!\!\lt[\! D_{\mu\nu}\!+\!2\eta_{\mu\nu}\psi \!+\!\fc{1}{2}\lt(\pt_{\mu}F_{\nu}\!+\!\pt_{\nu}F_{\mu} \rt)\!+\!2\pt_\mu\pt_\nu E \rt]\!\!,\label{hmunu}
\eeqn
where the vector modes satisfy the transverse condition $\pt^\mu S_{\mu}=\pt^\mu F_{\mu}=0$, and the tensor mode satisfies the transverse-traceless (TT) condition $\pt^\mu D_{\mu\nu}=0$. 

Correspondingly, the perturbed scalar fields are 
\beq
\phi^a=x^a+\pi^a,
\eeq 
where $\pi^a=\delta^a_\mu\pi^\mu$ is the Goldstone excitation of the condensation. In order to maintain the scalar condensation \eqref{Scalar_vacuum} under the general coordinate transformation $x^M \to x^M+\epsilon^M$, the Goldstone excitation has to transform opposite to the 4D coordinates simultaneously, i.e., $
\pi^\mu\to\pi^\mu-\epsilon^\mu$.  This St\"uckelberg trick non-linearly restores the general covariance of the theory. As a result, it behaves like a vector field and can be decomposed as $\pi^\mu=\eta^{\mu\nu}\lt(\pt_\nu \varphi+A_\nu \rt)$, with $\varphi$ a scalar field and $A_\mu$ a transverse vector field.

Consequently, there are 1 TT tensor, 3 transverse vector and 5 scalar modes in total. By decomposing $\epsilon_\mu =a^2(\epsilon^{V}_{\mu}+\pt_\mu \epsilon^s)$ with $\epsilon^{V}_{\mu}$ a transverse vector and $\epsilon^s$ a scalar, these 9 perturbed modes transform as follows under the general coordinate transformation,
\beqn
D_{\mu\nu} &\to& D_{\mu\nu}, \quad S_{\mu } \to S_{\mu } + a  \epsilon^{V}_{\mu }{}' ,\quad F_{\mu } \to  F_{\mu }-2 \epsilon^{V}_{\mu } ,\nn\\
A_{\mu } &\to&  A_{\mu }-{\epsilon }^{V}_{\mu }, \quad \psi \to \psi -H\epsilon _5,  \quad E \to E-\epsilon^s, \nn\\
\beta &\to& \beta+a {\epsilon }^s{}'+\frac{\epsilon _5}{a}, \quad  \xi \to \xi +\epsilon '_5, \quad \varphi \to \varphi -{\epsilon }^s.
\label{Gauge_transformation}
\eeqn

Naively, we have 5 gauge freedoms to eliminate one vector and two scalar modes by fixing ${\epsilon }^{V}_\mu$, ${\epsilon}^s$ and $\epsilon _5$. A commonly used gauge choice is the so-called unitary gauge, in which the Goldstone excitations $\pi^a$ of scalar fields are closed, i.e., $A_\mu=\varphi=0$. Moreover, we have another gauge freedom $\epsilon _5$ to set $\psi=0$. Then from the fact that $Z_\mu-\pi_\mu$, with $Z_\mu\equiv a^2\lt(F_\mu/2+\pt_\mu E \rt)$, is a gauge invariant quantity, one observes that the 4 Goldstone excitations  $\pi^\mu$ are ``eaten" by the 5D graviton in the unitary gauge. Consequently, the 5D massless spin-2 graviton with 5 DOF gets weight and becomes massive, with 9 DOF on the spectrum. After gauge fixing, we have  1 TT tensor, 2 transverse vector and 3 scalar modes left. However, not all modes among them are physical, since some of them can be eliminated by some constraint equations, which can be easily extracted by utilizing the Arnowitt-Deser-Misner decomposition. 

The Arnowitt-Deser-Misner decomposition of the 5-dimensional spacetime metric reads \cite{Manrique2011}
\beq
ds^2=N^2dy^2+\gamma _{\mu \nu } \left(dx^{\mu }+N^{\mu }dy\right) \left(dx^{\nu }+N^{\nu }dy \right),
\eeq
where $N$ is the lapse, $N^\mu$ the shift vector, and $\gamma _{\mu \nu }$ the induced metric on the slices. Correspondingly, the bulk action of \eqref{Main_Action} is rewritten as 
\beqn
S&=&\frac{M^3}{2}\int d^5x\sqrt{-\gamma }\Big[N \lt(R^{(4)}-2 \Lambda -m^2 \gamma ^{\mu \nu } \partial_\mu\phi ^a \partial_\nu\phi ^a \rt) \nn\\
&&-m^2 N^{-1}\left(\phi^{a}{}'-N^{\mu }\partial_{\mu }\phi ^a\right)\left(\phi^{a}{}'-N^{\nu }\partial_{\nu }\phi ^a\right)-N^{-1}\lt(E_{\mu \nu } E^{\mu \nu }-E^2\rt)\Big], \label{ADM_action}
\eeqn
where $E_{\mu \nu }=\frac{1}{2} \left(\gamma '_{\mu \nu }-\nabla _{\nu }N_{\mu }-\nabla _{\mu }N_{\nu }\right)$. The constraint equations are yielded by varying with respect to the lapse function $N$ and shift vector $N_\mu$, i.e.,
\beqn
R^{(4)}-2 \Lambda-m^2 \gamma ^{\mu \nu } \partial_\mu\phi^a \partial_\nu\phi^a +N^{-2}\big(E_{\mu \nu } E^{\mu \nu }\!&-&\!E^2\big)\nn\\
+m^2N^{-2}\left(\phi^{a}{}'-N^{\mu }\partial _{\mu }\phi ^a\right)\left(\phi^{a}{}'-N^{\nu }\partial_{\nu }\phi ^a\right)&=&0, \label{Constraint_Eq_I}\\
\nabla _{\nu }\lt(\fc{E^{\mu}_{\nu }-\delta^{\mu}_{\nu}E}{N}\rt)-\fc{m^2}{N}\partial_{\mu }\phi ^a\lt(\phi^a{}'-N^{\alpha } \partial_\alpha\phi^a\rt)&=&0. \label{Constraint_Eq_II}
\eeqn

Further, by considering small perturbations of the metric functions, expressed as $N=\bar N+\delta N$, $N^\mu=\bar N^\mu +\delta N^\mu$, and $\gamma _{\mu \nu }=\bar\gamma _{\mu \nu }+\delta \gamma _{\mu \nu }$, where $\bar N$, $\bar N^\mu$, and $\bar \gamma_{\mu \nu }$ represent the background quantities with $\bar N=1$, $\bar N^\mu=0$, and $\bar \gamma_{\mu \nu }=a^2(y)\eta_{\mu\nu}$, respectively. The perturbations $\delta N$, $\delta N^\mu$, and $\delta \gamma _{\mu \nu }$ can be explicitly extracted from Eqs.~\eqref{h55}, \eqref{hmu5}, and \eqref{hmunu}. Consequently, one can obtain a constraint equation for vector modes and two constraint equations for scalar modes. Thus, one vector mode and two scalar modes can be worked out algebraically, leaving us one vector and one scalar modes eventually.  

After scalar-vector-tensor decomposition and expanding the action \eqref{Main_Action} to the quadratic order of perturbations, the TT tensor, transverse vector and scalar modes are decoupled with each other, so they can be treated separately.

\subsection{Tensor mode} 
By including only the tensor perturbation in the perturbed metric and dropping the boundary terms, the quadratic action for the tensor perturbation reads
\beqn
S^{(2)}_T&=&\frac{M^3}{8}\int d^4xdya^2\lt[-{a^{2}} D_{ \alpha \beta }' D{}'^{\alpha \beta } -{\partial_\lambda D_{\alpha \beta }}\partial ^{\lambda }D ^{\alpha \beta }-2 m^2 D_{\alpha \beta } D^{\alpha \beta } \rt],
\eeqn
where the indices are raised and lowered by the 4D Minkowski metric $\eta_{\mu\nu}$. Further, after a coordinate transformation $dy=adz$ that turns the background metric into a conformal form and a rescaling $\tilde{D }={D }/{2}$, a canonical form is obtained as
\beqn
S^{(2)}_T&=&\frac{M^3}{2}\int d^4x dz a^3 \left[-\dot{\tilde{D }}_{\alpha \beta }\dot{\tilde{D }}^{\alpha \beta }-\partial_\lambda \tilde D_{\alpha \beta } \partial ^{\lambda }\tilde D ^{\alpha \beta }-2 m^2\tilde D_{\alpha \beta}  \tilde D^{\alpha \beta } \right],
\eeqn
where the dot denotes the derivative with respect to the extra dimension coordinate $z$. The tensor mode is free from the ghost instability due to the correct sign of kinetic term. Moreover, the action is expressed more concisely in the momentum space, where the d'Alembert operator $\partial ^{\alpha }\partial_{\alpha } $ is replaced by $ -p^2\equiv -p^\alpha p_\alpha$, with $p^\alpha$ the four-momentum of tensor mode. Then, it yields
\beq
S^{(2)}_T\!=\!\frac{M^3}{2}\!\!\int \!\!d^4 p dz a^3 \left[\!-\dot{\tilde{D }}_{\alpha \beta }\dot{\tilde{D }}^{\alpha \beta }\!-\!\left({p^2}\!+\!{2 m^2}\right)\tilde D_{\alpha \beta }\tilde D^{\alpha \beta } \right]\!\!.
\eeq
By variation with respect to $\tilde{D }^{\alpha \beta }$, we have the equation of motion
\beq
\ddot{\tilde{D} }_{\alpha\beta }+3H\dot{\tilde{D} }_{\alpha\beta }=(p^2+2m^2){\tilde{D}_{\alpha\beta } }.
\label{eom_tensor}
\eeq
Note that here and in what follows, $H \equiv \dot{a}/a$ refers to the extra dimension coordinate $z$.
After redefining $\tilde{D}_{\alpha\beta } = a^{-\fc{3}{2}}\mathcal{D}_{\alpha\beta }$, a Schr\"odinger-like equation is obtained,
\beq
-\ddot{\mathcal{D} }_{\alpha\beta } +\left(\frac{3}{2}\dot H+\frac{9}{4}H^2\right) {\mathcal{D} }_{\alpha\beta } =M_T^2{\mathcal{D}}_{\alpha\beta } ,
\label{SE_Tensor}
\eeq
where $M_T^2 \equiv -p^2-2m^2$. The Hamiltonian can be further factorized into a supersymmetric quantum mechanics form, with $H_T=A_T^\dag A_T=\lt(\pt_z+\fc{3}{2}H\rt)\lt(-\pt_z+\fc{3}{2}H\rt)$. Then, with the boundary condition $\pt_z \tilde{D}_{ \alpha \beta } |_{z=0,z_b} =0$, the self-adjoint Hamiltonian gives non-negative eigenvalues \cite{Yang2017}, i.e., $M_T^2\geq 0$. Consequently, with a positive 5D graviton mass $m$, the four-momentum of tensor mode $p^2=-M_T^2-2m^2<0$, i.e., the tensor excitations are all time-like particles. Thus, the model is also irrelevant to tachyonic instability. If $m$ is set to zero, $M_T^2=-p^2\geq 0$ gives us the well-known result that there is a massless graviton and a tower of massive gravitons in RS1 model.

\subsection{Vector modes} 

By including only the vector perturbations in the perturbed metric and dropping the boundary terms, the quadratic action for vector perturbations reads
\beqn
S_{V}^{(2)}\!&=&\!\frac{M^3}{16}\!\int \!d^4xdy a^2\Big[\!\!-\!a^2\pt_\beta F_\alpha{}'\pt^\beta F^\alpha{}'\!-\!2m^2\pt_\beta F_\alpha \pt^\beta F^\alpha -8a^2m^2 A_\alpha{}' A^\alpha{}'-8m^2 \pt_\beta A_\alpha \pt^\beta A^\alpha\nn\\
&&-4\partial_{\beta }S_{\alpha }\partial^\beta S^{\alpha }-8m^2S_\alpha S^\alpha+8m^2 \pt_\beta F_\alpha \pt^\beta A^\alpha -4a\pt_\beta F_\alpha{}'\pt^\beta S^\alpha-16 a m^2 S_\alpha A^\alpha{}'\Big].
\label{Action_SV}
\eeqn
The constraint equation can be obtained by counting the first order perturbations of Eq.~\eqref{Constraint_Eq_II}, or simplify by varying the above quadratic action with respect to $S^\alpha$, i.e.,
\beq
2\partial _{\beta }{\partial^\beta S_{\alpha }}-4 m^2 S_{\alpha }+a \partial_{\beta }{\partial^\beta  F_{\alpha }'}-4am^2A_\alpha'=0.
\label{Vector_Constraint_Eq}
\eeq
Then, $S^{\alpha }$ can be worked out in momentum space, namely,
\beq
S_{\alpha }=-\frac{a p^2 F_{\alpha }' +4 a m^2 A_\alpha '}{2 (p^2+2 m^2)}. 
\label{Vector_Constraint}
\eeq
Working in the unitary gauge, $A_\alpha$ is gauged away. Then, substituting the relation into the action \eqref{Action_SV} yields
\beq
S_V^{(2)}\!=\! M^3\!\int \!d^4 p dy  \!\left[-\frac{a^4 p^2 m^2 F_{\alpha }{}' F^{\alpha }{}'}{8 \left(p^2+2 m^2\right)}\!-\!\frac{a^2 p^2 m^2}{8} F_{\alpha } F^{\alpha } \right]\!.
\eeq
After performing a coordinate transformation into the $z$ coordinate, a canonical normalized form can be obtained by redefining the vector perturbation as
\beq
\tilde{F}_{\alpha }=\sqrt{\fc{p^2 m^2}{p^2+2 m^2}}\frac{ F_{\alpha }} {2},
\label{Redefined_Vector}
\eeq
yielding,
\beq
S^{(2)}_V\!=\!\frac{M^3}{2}\int d^4 p dz a^3  \left[-\dot{\tilde{F }}_{\alpha} \dot{\tilde{F }}^{\alpha}-\left({p^2}+{2 m^2}\right)\tilde{F}_{\alpha } \tilde{F}^{\alpha }\right]\!.
\eeq
The correct sign of the kinetic term ensures that the vector perturbation is free from the ghost instability. Then the  equation of motion reads
\beq
\ddot{\tilde{F} }_{\alpha }+3H\dot{\tilde{F} }_{\alpha }=(p^2+2m^2){\tilde{F}_{\alpha } }.
\eeq
A Schr\"odinger-like equation is given by redefining $\tilde{F}_{\alpha }\to a^{-\frac{3}{2}} \mathcal{F}_{\alpha }$,
\beq
-\ddot{\mathcal{F}}_\alpha+\left(\frac{3}{2}\dot H+\frac{9}{4}H^2\right) \mathcal{F}_\alpha=M_V^2\mathcal{F}_\alpha,
\label{SE_Vector}
\eeq
where $M_V^2 \equiv -p^2-2m^2$. The Hamiltonian can also be factorized into a supersymmetric quantum mechanics form, with $H_V=A_V^\dag A_V=\lt(\pt_z+\fc{3}{2}H\rt)\lt(-\pt_z+\fc{3}{2}H\rt)$. With the boundary condition $\pt_z \tilde{F }_{\alpha} |_{z=0,z_b} =0$, the eigenvalues are non-negative $M_V^2\geq 0$, i.e., $p^2 \leq -2m^2$. Thus, the vector excitations are also time-like particles and irrelevant to tachyonic instabilities. 

For the cases of $p^2 <-2m^2$,  the formula \eqref{Vector_Constraint} and the redefinition \eqref{Redefined_Vector} are valid. However, for the case of $M_V^2=-p_0^2-2m^2=0$,  the formula \eqref{Vector_Constraint} is problematic, and the constraint equation \eqref{Vector_Constraint_Eq} leads to 
\beq
F_\alpha'=2A_\alpha'.
\eeq
This implies that $F_\alpha-2A_\alpha \equiv f_\alpha(x)$ is a purely 4D field. Especially, the field $f_\alpha(x)$ is a gauge invariant quantity under the general coordinate transformation by the observation from \eqref{Gauge_transformation}. Then, the quadratic action \eqref{Action_SV} reduces to
\beqn
S^{(2)}_{V0} = -\frac{M^3}{8}\int d^4p dz a^3 m^2 k_0^2 \left(F_{\alpha }-2 A_{\alpha }\right) \left(F^{\alpha }-2 A^{\alpha }\right)=-\frac{M^3}{8}\int d^4 p dz a^3 m^2 k_0^2 f_\alpha f^\alpha.
\eeqn
The equation of motion reads $p_0^2f_\alpha=0$. Since $p_0^2=-2m^2$, it leads to $f_\alpha=0$. This implies that the lightest vector mode does not exist in the mass spectrum. This is curial to recover the mass spectrum of RS1 model in the massless limit $m=0$, where no massless vector mode exists due to the lack of continuous isometries of the higher dimension in the presence of 3-branes \cite{Randall1999}. Moreover, all the massive vector modes are gauge dependent in the massless limit $m=0$, therefore, they can be gauged away by gauge fixing.

\subsection{Scalar modes} 
By including only the scalar perturbations in the perturbed metric and dropping the boundary terms, the quadratic action for scalar perturbations reads
\beqn
S^{(2)}_{S}\!&\!=\!&\!M^3\!\!\int \!\!d^4xdy a^2\!\Big(6 a^2 H^2 \xi ^2\!+\!4 m^2 \psi ^2 \!-\! 4 m^2 \xi \psi \!+\! 6 a^2 \psi '{}^2+12 a^2 H \xi  \psi ' \!+\!3\pt_\alpha \psi \pt^\alpha \psi \!-\! \fc{m^2}{2} \pt_\alpha \beta \pt^\alpha \beta \!\nn\\
\!&&\! +\!3\xi \pt_\alpha \pt^\alpha \psi +3 aH\xi \pt_\alpha \pt^\alpha \beta + m^2 \xi \pt_\alpha \pt^\alpha\varphi-m^2 \xi \pt_\alpha \pt^\alpha E - 2m^2 \psi \pt_\alpha \pt^\alpha \varphi+2m^2\psi \pt_\alpha \pt^\alpha E \nn\\
\!&&\! +3 a^2 H \xi \pt_\alpha \pt^\alpha E' +3 a\beta \pt_\alpha \pt^\alpha\psi' + m^2 a \beta \pt_\alpha \pt^\alpha \varphi' -\fc{m^2}{2}  a^2 \pt_\alpha \varphi' \pt^\alpha \varphi'+3a^2 \psi'\pt_\alpha \pt^\alpha E' \nn\\
\!&&\! -\fc{m^2}{2}  \pt_\alpha \pt^\alpha E \pt_\lambda \pt^\lambda E -\fc{m^2 }{2} \pt_\alpha \pt^\alpha \varphi \pt_\lambda \pt^\lambda \varphi +m^2 \pt_\alpha \pt^\alpha \varphi \pt_\lambda \pt^\lambda E \Big).
\label{Scalar_full_action}
\eeqn
From the first order perturbation of Eqs.~\eqref{Constraint_Eq_I} and \eqref{Constraint_Eq_II} or simply varying the above action respect to the modes $\beta$ and $\xi$ respectively, the constraint equations are obtained as 
\beqn
12 a^2 H^2 \xi-4m^2\psi+12 a^2 H \psi'+3 \pt_\alpha \pt^\alpha \psi +3aH\!\!\!\!\!&\!\!\!\!\!&\!\!\!\partial _{\alpha }\partial ^{\alpha }\beta  \nn\\
+m^2 \partial _{\alpha }\partial ^{\alpha }\varphi
-m^2\partial _{\alpha }\partial ^{\alpha }E +3a^2H\partial _{\alpha }\partial ^{\alpha }E'&=&0,\\
m^2\partial_\alpha  \beta + 3aH\partial _{\alpha }\xi + 3a \pt_\alpha   \psi' +m^2 a \pt_\alpha  \varphi' &=&0.
\eeqn
After closing the scalar perturbations $\varphi$ and $\psi$ in the unitary gauge, we only have to take the remaining perturbations $\xi$, $\beta$ and $E$ into account in the quadratic action, i.e.,  
\beqn
S^{(2)}_{S}&=&M^3\int d^4xdy a^2\Big(6 a^2 H^2 \xi ^2 -\frac{m^2}{2}\partial _\alpha \beta \partial ^{\alpha }\beta+3 a H \xi\partial _{\alpha }\partial ^{\alpha }\beta-m^2\xi \partial _{\alpha }\partial ^{\alpha }E\nn\\
&&+3a^2H\xi \partial _{\alpha }\partial ^{\alpha }E'-\frac{m^2}{2}\partial _{\alpha }\partial ^{\alpha }E\partial _{\lambda }\partial ^{\lambda }E\Big).
\label{Scalar_action}
\eeqn
Correspondingly, the constraint equations are rewritten as 
\beqn
12 a^2 \!H^2 \xi\!+\!3aH\partial _{\alpha }\partial ^{\alpha }\!\beta \!-\!m^2\partial _{\alpha }\partial ^{\alpha }\!E \!+\!3a^2\!H\partial _{\alpha }\partial ^{\alpha }\!E'\!&\!=\!&0,\\
m^2\partial_\alpha  \beta+3aH\partial _{\alpha }\xi  \!&\!=\!&0.
\eeqn
In momentum space,  $\beta$ and $\xi $ can be worked out from the constraint equations as 
\beqn
\beta &=&-\frac{3 aH}{m^2}\xi ,\\
\xi &=&\frac{p^2m^2(3 a^2 HE'- m^2 E)}{3 a^2 H^2 \left(3 p^2+4 m^2\right)}.
\eeqn
Thus, the perturbations $\beta$ and $\xi$ can be eliminated by substituting these relations into the action \eqref{Scalar_action}, then a quadratic action for scalar perturbation $E$ is achieved as
\beq
S^{(2)}_{S}\!\supset \!M^3\!\! \int \!\!d^4 p dy \fc{3a^2 p^4 m^2}{6 p^2\!+\!8 m^2}  \left[-a^2E' E' \!-\!\lt(p^2\!+\!2 m^2\rt)\!E^2\right]\!.
\eeq
Further, after redefining the scalar perturbation as 
\beq
\tilde E= \sqrt{\frac{-3 p^4 m^2}{3 p^2+4 m^2}}E,
\label{Redefined_Scalar}
\eeq
 the above action can be rewritten as a canonical normalized form in the conformal coordinate $z$, i.e.,
\beq
S^{(2)}_{S}\supset\frac{M^3}{2}\int d^4 p dz a^3 \left[\dot{\tilde{E} }\dot{\tilde{E}} +\left({p^2}+{2 m^2}\right)\tilde{E}^2\right].
\eeq 
It is clear that there is an overall wrong-sign in the Lagrangian, which implies that the scalar mode $\tilde E$ is a ghost field. However, the equation of motion is not affected by the overall sign of the Lagrangian, so it has a similar form to the tensor and vector modes,  given by
\beq
\ddot{\tilde{E} }+3H\dot{\tilde{E} }=(p^2+2m^2){\tilde{E} }.
\eeq
By redefining $\tilde{E}=a^{-\fc{3}{2}}{\varepsilon}$, it can be rewritten as  a Schr\"odinger-like equation,
\beq
-\ddot{{\varepsilon }}+\left(\frac{9 }{4}H^2+\frac{3 }{2}\dot H\right){\varepsilon } = M_S^2{\varepsilon},
\label{SE_Scalar}
\eeq
where $M_S^2\equiv-p^2-2m^2$. The Hamiltonian can be factorized as a supersymmetric quantum mechanics form as well, $H_S=A_S^\dag A_S=\lt(\pt_z+\fc{3}{2}H\rt)\lt(-\pt_z+\fc{3}{2}H\rt)$, so with the boundary condition $\pt_z \tilde E |_{z=0,z_b} =0$, the eigenvalues are non-negative $M_S^2\geq 0$, i.e., $p^2\leq-2m^2$. Thus, the redefinition \eqref{Redefined_Scalar} is well-defined. In this case, all the KK excitations of the scalar mode are time-like particles.

In RS1 model, the presence of IR brane abruptly ends AdS space, so it spontaneously breaks the conformal invariance of AdS bulk in the IR. The massless radion is just the Goldstone boson associated with the broken dilatation invariance \cite{Arkani-Hamed2001a,Rattazzi2001}. For the current model, the scalar curvature reads $R=-20k^2+\fc{4m^2e^{2k y}}{1+\epsilon^2 e^{2ky}}$ in the bulk. Thus, the presence of scalar condensation induces a slight deformation in the bulk geometry, causing it to deviate from a pure AdS space. Consequently, the radion acquires a tiny mass due to this explicit symmetry breaking. In the massless limit $m=0$, the quadratic action \eqref{Scalar_full_action} will reduce to that of RS model, with only a massless radion in the mass spectrum \cite{Goldberger1999a,Charmousis2000,Callin2005}.

\section{Mass Spectra and Hierarchy Resolution}\label{Hierarchy}

For the lightest tensor and scalar modes corresponding to $M_{T,S}=0$, their wave functions can be easily solved from Eqs.~\eqref{SE_Tensor} and \eqref{SE_Scalar}. After returning to the coordinate space,  and utilizing the KK decompositions, $\mathcal{D}_{\alpha\beta }(x,z) =d_{\alpha\beta }(x)\Psi(z)$ and $\varepsilon(x,z) =e(x)\Psi(z)$,   a Schr\"odinger-like equation
\beq
-\ddot{\Psi} +\left(\frac{3}{2} \dot{H}+\frac{9}{4}H^2\right)\Psi =M_{T,S}^2\Psi,
\eeq
with $M_{T,S}^2=m_{T,S}^2-2m^2$, can be achieved from Eqs.~\eqref{SE_Tensor} and \eqref{SE_Scalar}. Here, $m_{T,S}$ is the effective mass of KK particles in 4D point of view, satisfying the 4D Klein-Gordon equations $\Box^{(4)}d_{\mu\nu}=m_T^2 d_{\mu\nu}$ and $\Box^{(4)}e(x)=m_S^2 e(x)$. 

The wave function of ground states is easily achieved by setting $M_{T,S}=0$ or $m_{T,S}=\sqrt{2}m$ in above Schr\"odinger-like equation, i.e.,
\beq
\Psi^{(0)}(z)=N_0a(z)^{\fc{3}{2}},
\eeq 
where $N_0$ is the normalization factor.  The boundary conditions $\pt_z \tilde{D}_{ \alpha \beta } |_{z=0,z_b} =0$ and $\pt_z \tilde E |_{z=0,z_b} =0$ lead to $\lt.\lt( \Psi-\frac{3}{2}H \rt)\rt|_{z=0,z_b}=0$. Obviously, the wave function $\Psi^{(0)}(z)$ satisfies the boundary condition. All KK particles are massive in this model, which is a significant difference from the RS1 model, where the lightest KK particles are massless spin-2 graviton and spin-0 radion. Especially, from the redefinitions $\tilde{D}_{\alpha\beta } = a^{-\fc{3}{2}}\mathcal{D}_{\alpha\beta }$ and $\tilde{E}=a^{-\fc{3}{2}}{\varepsilon}$, their canonical normalized field configurations are given by $\tilde{D}^{(0)}_{\alpha\beta }=a^{-\fc{3}{2}}\mathcal{D}^{(0)}_{\alpha\beta }=d^{(0)}_{\alpha\beta }(x)$ and $\tilde{E}^{(0)}=a^{-\fc{3}{2}} {\varepsilon}^{(0)} =e^{(0)}(x)$. Therefore, the lightest graviton and radion propagate only on the brane. 

However, the effective mass of 4D graviton is severely constrained by the gravitational experiments \cite{Rham2017}. For instance, the detection of gravitational waves has provided a bound of the graviton mass, yielding $m_g\leq 4.7\times 10^{-23}$eV \cite{Abbott2019}. Thus, the parameter $m$ must be tinier than the experimental constraints, i.e., $m<3.3\times 10^{-23}$eV. Since the lightweight radion has the same mass as the lightest graviton, exchanging such nearly massless scalar particle would cause a fifth force and violate experimental observations. Therefore, similar to the RS1 model, the radion must gain weight to meet the experimental expectations, which is realized through GW mechanism \cite{Goldberger1999a} in the next section.

With the normalization condition $\int^{z_b}_{-z_b}\Psi_0^2dz=1$, the normalization factor is worked out as 
\beqn
N_0^{-2}=\frac{1}{k}\lt[1-e^{-2 k y_\pi}+4\epsilon^2 k y_\pi -\epsilon^4(1-e^{2 k y_\pi})\rt].
\label{Normalization_factor}
\eeqn
By shutting down the 5D graviton mass $m$, one recovers the result of RS1 model. However, if we remove the visible brane, i.e., $y_\pi\to\infty$, the quasi-massless graviton is no longer normalizable. This means that the effective 4D gravity theory cannot be recovered on the brane. Therefore, the RS2-like non-compact single brane model \cite{Randall1999a} is not a physically available one in current massive gravity.

The braneworld scenario provides a natural way to solve the  gauge hierarchy problem, which is a crucial motivation of the well-known ADD model \cite{Arkani-Hamed1998} and RS1 model \cite{Randall1999}. In our toy model,  the low-energy effective  theory is obtained by including the nearly-massless gravitons, i.e.,
\beqn
ds^2&=&a^2(y)\bar g_{\mu\nu}(x)dx^\mu dx^\nu+dy^2\nn\\
&=&a^2(y)\lt[\eta_{\mu\nu}+\gamma_{\mu\nu}(x)\rt]dx^\mu dx^\nu+dy^2,
\label{Metric_Lowenergy}
\eeqn
then the 4D effective gravitational mass scale $M_{\text{eff}}$ is read from the curvature term in the action \eqref{Main_Action},
\beq
M_{\text{eff}}^2=M^3\int^{y_\pi}_{-y_\pi} a^2 dy={N_0^{-2}}{M^3}. \label{Mass_scale_relation}
\eeq

On the other hand, in order to produce a large hierarchy between the Planck scale and the electroweak scale, our Universe should be confined to the IR brane located at $y_\pi$. Then, the Higgs field action on the brane is non-canonically normalized, 
\beq
S_\mathcal{H}\supset \! \int{  d^4x \sqrt{| g_\text{II}|}\lt[-{g}_\text{II}^{\mu\nu}D_{\mu}\mathcal{H}^{\dag}D_{\nu}\mathcal{H}-\lambda(\mathcal{H}^{\dag}\mathcal{H}-v_0^2)^2 \rt]  },
\eeq
where $g_{\text{II}\mu\nu}=a(y_\pi)\tilde{g}_{\mu\nu}$ is the induced metric on the brane at $y_\pi$, and $v_0$ the fundamental Higgs vacuum expectation value (VEV). Writting the warp factor explicitly, one has
\beqn
S_\mathcal{H}&\supset&\int  d^4x \sqrt{|\tilde g|}\lt[-a(y_\pi)^{2}\tilde{g}^{\mu\nu}D_{\mu}\mathcal{H}^{\dag}D_{\nu}\mathcal{H}-a(y_\pi)^4\lambda(\mathcal{H}^{\dag}\mathcal{H}-v_0^2)^2 \rt],
\eeqn
 With a field renormalization, $\mathcal{H} \to \tilde{\mathcal{H}}/a(y_\pi)$, the effective action of Higgs on the brane is 
 \beq
 S_\mathcal{H}\!\supset\! \int{  d^4x \sqrt{|\tilde g|}\lt[-\tilde{g}^{\mu\nu}D_{\mu}\tilde{\mathcal{H}}^{\dag}D_{\nu}\mathcal{H}-\lambda(\tilde{\mathcal{H}}^{\dag}\mathcal{H}-v_\text{eff}^2)^2 \rt]  },
 \eeq
 where the effective Higgs VEV, $v_\text{eff}=a(y_\pi) v_0$, sets the electroweak scale on the brane.  
 
Therefore, in order to solve the gauge hierarchy problem, the fundamental parameters $M$, $k$, $v_0$ are simply chosen to be of the same order as the Planck scale, $M_\text{Pl}$. Then, the warp factor has to provide enough redshift to recover a TeV scale of Higgs VEV on the brane, namely, $a(y_\pi)\sim10^{-16}$. In the condition $e^{-k |y_\pi|}>\epsilon$ and $m<3.3\times 10^{-23}$eV, it leads to $e^{-ky_\pi} \approx a(y_\pi) -\fc{\epsilon^2}{a(y_\pi)} \sim10^{-16}$. Thus, it requires  $y_\pi\approx 37/k$.

Since the mass splitting scale of massive KK modes is inversely proportional to the conformal size of extra dimension $z_b$, i.e., $\Delta m_T \propto 1/z_b$. From the coordinate transformation $dy=adz$, one has
\beq
z_\pi=\frac{1}{k \epsilon} \left[\text{arctan}\left(\epsilon e^{k y_\pi}\right)-\text{arctan} \left(\epsilon \right)\right],
\eeq
where the integral constant has been chosen so that $z(y=0)=0$. Since $e^{-ky_\pi}  \sim10^{-16}$, $m<3.3\times 10^{-23}$eV and $k\sim M_\text{Pl}$, one has $\epsilon e^{k y_\pi} \ll 1$. Thus, the formula can be rewritten approximately as $z_\pi \approx \fc{e^{ky_\pi}}{k}\lt(1+\mathcal{O}(\epsilon^2)\rt)$. So the mass splitting scale is similar to the RS1 model, i.e., $\Delta m_T \sim  k e^{-ky_\pi} \sim \mathcal{O}(\text{TeV})$.

\section{Radius stabilization}\label{Radius_stabilization}

In RS1 model, the position of IR brane is put by hand, namely, the radius is not dynamically fixed. It leads to the presence of a massless radion in the effective theory, which corresponds to the fluctuations of the radius of compact extra dimensions. However, the massless radion is phenomenologically unacceptable since it would contribute to Newton's law and cause a fifth force. It is well-known that the GW mechanism \cite{Goldberger1999a} can be introduced to stabilize the size of the extra dimension and to increase the mass of radion. In current model, the position of IR brane is put by hand as well, we can also utilize the GW mechanism here to stabilize the brane position and increase the radion mass. 

By adding a bulk scalar field $\Phi$ into the model, which has interaction with the two branes, the action is given by
\beqn
S_\Phi&=&\fc{1}{2}\int d^4x\int^\pi_{-\pi}d\theta\sqrt{-g}\lt(-g^{MN}\pt_M\Phi\pt_N\Phi-m_\Phi^2\Phi^2 \rt)-\int{}d^4x\sqrt{-g_\text{I}}\lambda_\text{I}(\Phi^2-v_\text{I}^2)^2\nn\\
&&-\int{}d^4x\sqrt{-g_{\text{II}}}\lambda_{\text{II}}(\Phi^2-v_{\text{II}}^2)^2,
\label{GW_action}     
\eeqn
where $g_{MN}$ is the background metric \eqref{Brane_Metric} with the radius $r_c$ of compactified extra dimension given by $y=r_c \theta$, $\lambda_{\text{I/II}}$ is the coupling parameters, and $V_{\text{I/II}}=\lambda_{\text{I/II}} v_{\text{I/II}}^4$.

The equation of motion of the scalar filed $\Phi$ is achieved by varying the action \eqref{GW_action} with respect to  $\Phi$, 
\beqn
\frac{1}{r_c}{\partial_\theta }\left(a^4 {\partial_\theta  \Phi }\right)-a^4 m_{\Phi }^2\Phi-{4 a^4  \lambda_\text{I} \Phi  \left(\Phi ^2-v_\text{I}^2\right)}\delta (\theta)
-4 a^4\lambda _\text{II} \left(\Phi ^2-v_\text{II}^2\right) \Phi  \delta (\theta -\pi ) =0.&&
\label{EoM_Phi}
\eeqn
Away from the two branes at $\theta=0,\pi$, the general solution of this equation reads
\beqn
\Phi(\theta)=e^{2 \sigma } \lt[A  {}_2\text{F}_1\left(2,\nu +2,\nu +1, -\epsilon ^2e^{2 \sigma} \right) e^{ \nu \sigma }+B {}_2\text{F}_1\lt(2,2-\nu ,1-\nu ,-\epsilon ^2e^{2 \sigma } \rt) e^{-\nu \sigma }\rt],
\eeqn
where $\sigma(\theta)=k r_c \theta$, $\nu=\sqrt{4+m_\Phi^2/k^2}$, ${}_2\text{F}_1$ is the hypergeometric function, $A$ and $B$ are integration constants. Especially, under the condition $\epsilon e^{\sigma} \ll 1$, up to the first order of correction, the solution can be approximately written as
\beqn
\Phi(\theta)\simeq e^{2 \sigma } \lt[A  \left(1-2 \epsilon^2\frac{\nu +2}{\nu +1}e^{2 \sigma } \right)e^{ \nu \sigma }+B \left(1-2\epsilon^2\frac{\nu -2}{\nu -1} e^{2 \sigma }\right)e^{ -\nu \sigma }\rt].
\eeqn
 If we close the mass of 5D graviton, i.e., $\epsilon=0$, the solution reduces to the one in GR \cite{Goldberger1999a}. The boundary conditions on the branes can be obtained by inserting the approximate solution into the equation of motion \eqref{EoM_Phi} and matching the delta functions, given by
 \beqn
 k\! \left[ (\nu \!+\!2) \!\left(1\!+\!2\epsilon ^2\frac{\nu -2}{\nu +1}\right)\!A\!-\! (\nu\!-\!2 ) \!\left(1\!+\!2\epsilon ^2\frac{\nu+2}{\nu -1}\right)\!B\right]\nn\\
 -2 \lambda _\text{I}   \left(1+4 \epsilon ^2\right)  \Phi(0)\left( \Phi(0)^2 -v_\text{I}^2\right)=0,~~~\\
 k e^{2  \sigma(\pi) } \left[(\nu +2) e^{\nu  \sigma(\pi)} \left(1+2\epsilon ^2\frac{\nu-2}{\nu +1} e^{2 \sigma(\pi)}\right)A -(\nu-2) e^{ - \nu \sigma(\pi) } \left(1+2\epsilon ^2\frac{\nu +2}{\nu -1} e^{2  \sigma(\pi) }\right) B \right]\nn\\
 +2 \lambda _v \left(1+4 \epsilon ^2 e^{2 \sigma(\pi)}\right)\Phi(\pi )\left(\Phi(\pi)^2 -v_v^2\right) =0.~~~
 \eeqn 
 The integration constants $A$ and $B$ can be determined from these boundary conditions. 
 
 However, rather than solving the above equations directly, we employ the trick of Ref.~\cite{Goldberger1999a} to simplify the calculation. By substituting the approximate solution back into the action and integrating over $\theta$, it yields the effective potential of compactified radius $r_c$, 
 \beqn
 V_\Phi(r_c)&\approx& k A^2  (\nu +2) e^{2 \nu k  r_c  \pi } \left(1-\frac{8 \epsilon^2 e^{2 \pi  k r_c}}{\nu +1}\right)+ e^{-4 k r_c \phi }\lambda _\text{II} \left(1+4 \epsilon ^2 e^{2 \pi  k r_c}\right)\left( \Phi(\pi)^2 -v_\text{II}^2\right)^2\nn\\
&&+\lambda _\text{I} \left( \Phi(0)^2 -v_\text{I}^2\right)^2,
\label{Potential_Phi}
 \eeqn
 where the limit of $e^{k r_c \pi}\gg 1$ has been used in the calculation. Assuming that the interaction parameters $\lambda_\text{I}$ and $\lambda_\text{II}$ are very large  \cite{Goldberger1999a}, this effective potential implies the solution  $\Phi(0)=v_\text{I}$ and $\Phi(\pi)=v_\text{II}$. Then, the integration constants $A$ and $B$ can be solved approximately, yielding 
 \beqn
 A&\approx & v_\text{II} e^{-(\nu +2) \sigma(\pi) } \left[1-2\epsilon^2\frac{\nu +2}{\nu +1} e^{2 \sigma(\pi) } \right]-v_\text{I} e^{-2 \nu  \sigma(\pi) } \left[1+4\epsilon^2\frac{ \nu }{\nu ^2-1} e^{2 \sigma(\pi) }\right],\\
 B&\approx & v_\text{I} \left[1+2\epsilon^2\frac{\nu +2}{\nu +1}  \left(e^{\nu  \sigma(\pi) }-e^{2 \sigma(\pi) }\right)\right]-v_\text{II} e^{-(\nu +2) \sigma(\pi) } \left[1+2\epsilon^2\frac{\nu +2}{\nu +1}  e^{\nu  \sigma(\pi) }\right].
 \eeqn
Further, assuming that the mass of the scalar filed is a small quantity, i.e., ${m_\Phi}/{k}\ll 1$, one has $\nu =\sqrt{4+\frac{m_\Phi^2}{k^2}}\simeq 2+\delta$, where $\delta=\frac{m_\Phi^2}{4 k^2}$ is a small quantity. With this approximation, the effective potential can be rewritten as
\beqn
V_\Phi(r_c)&\approx& (4+\delta) k e^{-4 k r_c \pi}\left(1-\frac{8 \epsilon ^2e^{2 k r_c \pi } }{\delta +3}\right)\left[\left(1+\frac{4 (\delta +2) \epsilon ^2e^{2 k r_c \pi } }{\delta ^2+4 \delta+3 }\right)e^{-\delta  k r_c \pi} v_\text{I} \rt.\nn\\
&&-\lt.\left(1-\frac{2 (\delta +4)  \epsilon ^2e^{2k r_c \pi }}{\delta +3}\right)v_\text{II} \right]^2.
\eeqn 
As a result, the effective potential has a minimum at 
\beq
r_c \approx \frac{1}{\pi k\delta }\ln\left(\frac{v_h}{v_v}\right)-\frac{4\epsilon ^2}{3\pi k}  \left(\frac{v_h}{v_v}\right)^{\fc{2}{\delta }}\left(1+\frac{v_v}{v_h}\right).
\eeq
The first leading term is just the result of GW mechanism in GR \cite{Goldberger1999a}, and the second term proportional to $\epsilon^2$ is a tiny correction stemming from the 5D graviton mass. The mechanism provides a dynamical way to stabilize the compactified radius of the extra dimension without introducing another large hierarchy. For instance, if $m_\Phi/k=0.2$ and ${v_h}/{v_v}=1.45$, one obtains $ky_\pi=k r_c\pi\approx 37$ to generate a proper hierarchy. After the radius stabilization, the radion acquires a mass roughly $\mathcal{O}(\delta)$TeV \cite{Goldberger2000}, which is still smaller than the TeV scale.

\section{Conclusions}\label{Conclusions}

In this work, we generalized the RS1 model in a 5D extension of the Lorentz-violating massive gravity. It is found that the theory supports two distinct brane configurations. The configuration possessing both positive and negative tension branes is similar to the RS1 model, while the other possessing only two positive tension branes is distinct from RS1 model. The gauge hierarchy problem was also solved for the  RS1-like brane configuration as an application.  In order to analyze the mass spectra and stability, we considered full linear perturbations against the background metric. It is found that the tensor and vector modes are free from the ghost and tachyonic instabilities, however, the scalar mode is a ghost filed. 

\begin{figure}[t]
\begin{center}
\includegraphics[width=8cm]{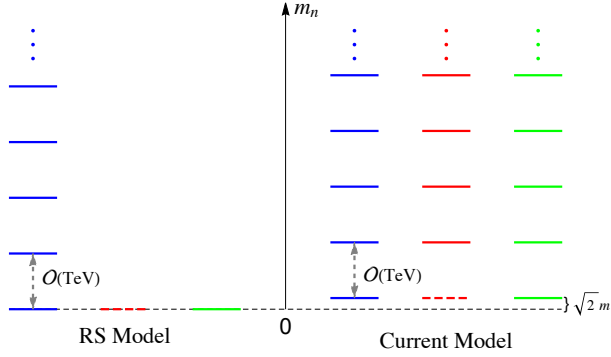}
\end{center}
\caption{The mass spectra of RS1 model and current model, where the blue lines refer to the tensor mode, red lines to vector mode, and green lines to ghost scalar mode.}
\label{Mass_Spectra}
\end{figure}

 As shown in Fig.~\ref{Mass_Spectra},  all the KK particles are massive. The tensor and scalar modes have similar mass spectra which start from $\sqrt{2}m$ with a mass splitting of TeV scale, nevertheless, the lightest vector mode does not exist in the mass spectrum. Furthermore, the ground states of tensor and scalar modes propagate only along the brane. The graviton mass $\sqrt{2}m$ has to be tiny enough to fit the experimental constraints. Consequently, the low-energy effective theory on the brane is a massive gravity plus a ghost scalar field. Although the ghost scalar filed would lead to quantum instability, it is interesting to note that the gravity coupled to a real ghost scalar field with a negative sign for the kinetic and potential terms can support non-singular, static solutions for topologically non-trivial wormhole-like geometry \cite{Kodama1978,Kodama1979}.  

Further, by imposing the GW mechanism to stabilize the size of the extra dimension, the light weight ghost radion would get weight to be the order of $\mathcal{O}(\delta)$ TeV.  After the radius stabilization, the 4D effective theory on the brane includes a nearly massless graviton plus three towers of very massive spin-2, spin-1 and ghost spin-0 particles.

In order to achieve a Minkowski flat brane model, the 4D Poincar\'e invariance is required to be preserved on the brane. Therefore, the internal metric of the scalar fields is chosen to be the Minkowski metric in the action \eqref{Main_Action}. However, the non-positive definite Minkowski metric leads to the bulk scalar $\phi^0$ having a kinetic term with the wrong sign. Since there is no further constraint to eliminate this ghost scalar mode, it results in the presence of the ghost in the linear theory. In order to get rid of the ghost scalar mode, we can choose the internal metric to be $\delta_{ij}$ with $i,j=1,2,3$, which is positive definite and removes the problematic mode $\phi^0$. However, in this scenario, the scalar condensation $\langle\phi^a\rangle=\delta_i^a x^i$ breaks the 4D Lorentz symmetry and preserves the $SO(3)$ symmetry on the brane. As a result, we cannot achieve a Minkowski flat brane model anymore. Instead, the brane needs to be bent, for example, de Sitter brane. The attempt to establish such a stable bent brane model is left for our another work.

\section*{ACKNOWLEDGMENTS}

We would like to especially thank Prof.~Yu-Xiao Liu for a very helpful discussion of our paper. This work was supported by the National Natural Science Foundation of China under Grant Nos.~12005174 and 12165013. K. Yang acknowledges the support of Natural Science Foundation of Chongqing, China under Grant No. cstc2020jcyj-msxmX0370. B.-M. Gu is also supported by Jiangxi Provincial Natural Science Foundation under grant No. 20224BAB211026.


\end{document}